\documentclass[twocolumn,aps,showpacs]{revtex4}
\usepackage[latin9]{inputenc}
\usepackage{textcomp}
\usepackage{amsmath}
\usepackage{amssymb}
\usepackage{graphicx}
\usepackage{setspace}
\usepackage[unicode=true,pdfusetitle,
 bookmarks=true,bookmarksnumbered=false,bookmarksopen=false,
 breaklinks=false,pdfborder={0 0 1},backref=section,colorlinks=false]
 {hyperref}
\usepackage{breakurl}

\makeatletter

\providecommand{\tabularnewline}{\\}

\@ifundefined{textcolor}{}
{%
 \definecolor{BLACK}{gray}{0}
 \definecolor{WHITE}{gray}{1}
 \definecolor{RED}{rgb}{1,0,0}
 \definecolor{GREEN}{rgb}{0,1,0}
 \definecolor{BLUE}{rgb}{0,0,1}
 \definecolor{CYAN}{cmyk}{1,0,0,0}
 \definecolor{MAGENTA}{cmyk}{0,1,0,0}
 \definecolor{YELLOW}{cmyk}{0,0,1,0}
}

\usepackage{color}%\bibliographystyle{plain}
\def\NOT(#1,#2){\OneQubitGate(#1,#2){$X$}}

\makeatother

\begin{document}

\title{High-fidelity gate operations for quantum computing beyond dephasing
time limits.}

\author{Alexandre M. Souza$^{1}$, Roberto S. Sarthour$^{1}$, Ivan S. Oliveira$^{1}$}

\affiliation{$^{1}$Centro Brasileiro de Pesquisas Físicas, Rua Dr. Xavier Sigaud
150, Rio de Janeiro 22290-180, RJ, Brazil}

\author{Dieter Suter$^{2}$}

\affiliation{$^{2}$Fakultät Physik, Technische Universität Dortmund, D-44221
Dortmund, Germany}
\begin{abstract}
The implementation of quantum gates with fidelities that exceed the
threshold for reliable quantum computing requires robust gates whose
performance is not limited by the precision of the available control
fields. The performance of these gates also should not be affected
by the noisy environment of the quantum register. Here we use randomized
benchmarking of quantum gate operations to compare the performance
of different families of gates that compensate errors in the control
field amplitudes and decouple the system from the environmental noise.
We obtain average fidelities of up to 99.8\%, which exceeds the threshold
value for some quantum error correction schemes as well as the expected
limit from the dephasing induced by the environment.
\end{abstract}

\pacs{03.67.Pp, 03.67.Lx}

\maketitle
\begin{singlespace}
Scalable quantum computing requires gate operations with fidelities
above a certain threshold \cite{2711,3921}. Reaching this threshold
remains challenging, primarily due to two issues: (i) the control
fields driving the gate operations have often experimental uncertainties,
such as amplitude errors, that are bigger than the allowed deviations
and (ii) the quantum systems cannot be completely shielded from the
effects of a noisy environment. Different techniques have been developed
to overcome these obstacles, such as dynamical decoupling (DD) for
suppressing the effect of environmental noise and thus extend the
coherence time of the system. An ideal dynamical decoupling sequence
completely eliminates the interactions of the system with its environment
and thereby ``freezes'' the system, apart from the evolution under
the internal system Hamiltonian. This approach can therefore implement
high-fidelity quantum memories. If, however, the specific application
requires that the system evolves under a suitable control Hamiltonian,
such as in a quantum information processor, the protection scheme
and the control operations must be applied simultaneously. It is then
necessary to design gates that combine processing with decoupling
in such a way that the effect of the control fields survives, while
all unwanted interactions are eliminated \cite{PhysRevLett.112.050502}.
\end{singlespace}

Measuring the fidelity of gate operations in the region of the threshold
for scalable quantum computation cannot be done on individual gates,
since the measurement errors can be comparable to or greater than
the gate errors. For this purpose, the ``randomized benchmarking''
procedure was developed \cite{2713}. It measures an \emph{average}
gate fidelity of a random selection of gates. This allows a much more
precise determination of the average gate fidelity. It does not provide
information about the fidelity of individual gates, but it provides
estimates of the fidelity of sequences of gates, which is the relevant
quantity for the analysis of large-scale computation.

In this paper, we present several families of gate operations that
combine robustness against unwanted variations in the amplitudes of
the control fields (flip angle errors) with protection against environmental
noise. For this purpose, we use Clifford gates, which allow universal
quantum computation when combined with magic states \cite{PhysRevA.71.022316,Souza:2011fk,PhysRevA.91.022314}.
We determine their fidelities experimentally, using a nuclear spins
as qubit and a different nuclear spin system as the noisy environment. 

We consider a system qubit $S$, which is coupled to a noisy environment
via a dephasing interaction 
\begin{equation}
\mathcal{H}_{SE}=b(t)S_{z},\label{eq:HSE}
\end{equation}
where the (semi-)classical field $b(t)$ has a finite correlation
time $\tau_{c}$. Dynamical decoupling by sequences of inversion pulses
can suppress the dephasing effect of this interaction and prevent
the decay of the coherence in the system \cite{5916,PhysRevA.82.042306}.
This is well explored in the context of quantum memories, where the
goal is to keep a specific quantum state unchanged. Here the effect
of DD can be roughly described as a prolongation of the dephasing
time $T_{2}$.

In the case of quantum computing, the goal is not the preservation
of a quantum state, but the precise control of a quantum system such
that it evolves along a well-defined path in Hilbert space. Dephasing
then creates errors roughly $\propto(1-e^{-\tau/T_{2}})\approx\tau/T_{2}$,
where $\tau$ is the duration of the gate operation and $T_{2}$ the
dephasing time. Extending the dephasing time by DD therefore helps
also for quantum computing. However, in this case, an additional complication
arises: the dynamical decoupling sequences that are used for protecting
quantum memories would equally decouple the system from the external
fields that are applied to control the system's path. It is therefore
essential to combine dynamical decoupling and gate operations in such
a way that they do not interfere destructively. Two general approaches
for this have been proposed for solving this problem: to apply gate
operations and DD operations successively \cite{PhysRevA.84.012305},
or to interleave gate operation and DD by splitting the gate operation
into as many elements as there are delays between the DD pulses and
modify them in such a way that the effect of the DD pulses is to revert
these modifications and the overall effect becomes that of the targeted
gate operation \cite{khodjasteh:080501,PhysRevLett.104.090501,prl230503,PhysRevA.80.032314,PhysRevA.84.012305,PhysRevLett.102.210502,PhysRevLett.112.050502,PhysRevA.86.050301}.

A third option is to design gate operations that are robust against
experimental imperfections and have `built-in' the effect of a dynamical
decoupling sequence. Such a decoupling effect is, e.g., built into
$\pi$-pulses around an axis in the $xy$-plane: they invert the system
operator $S_{z}$ and therefore the system environment interaction
given in eq. (\ref{eq:HSE}). The sequences that we consider for randomized
benchmarking consist to 50\% of $\pi$-pulses. Accordingly, they automatically
reduce the effect of the system-environment interaction, although
their efficiency may be lower than that of specialized DD-sequences.

\begin{figure}
\centering{}\includegraphics[clip,scale=0.5]{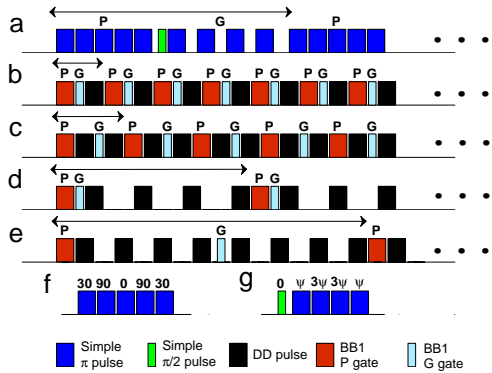}\protect\caption{Different sequences of gate operations compared in this work. The
double arrows indicate the gate duration $\tau$, defined as the time
period between the beginning of one P gate to the beginning of the
next P gate. a) A protected G rotation is created by separating in
time the four $\pi$-pulses of the BB1 composite rotation by a delay
that is twice the duration of the initial $\pi/2$-pulse, while the
P rotation was implemented by one KDD-5 pulse. From b) to e) PG gates
were implemented by BB1 pulses and interleaved in different ways with
the DD sequence XY-16 (see text). f) and g) show the KDD-5 and BB1
pulses. \label{fig:Different-versions-of} }
\end{figure}

Apart from reducing the coherence time, high fidelity operations also
need to be robust against errors in the control fields driving the
gate operations. The BB1 composite pulse (see figure \ref{fig:Different-versions-of}g)
was introduced by Wimperis \cite{Wimperis:1994fk} as a possible scheme
for generating composite rotation pulses that are well compensated
against amplitude errors. The sequence for generating a compensated
rotation by an angle $\theta$ around an axis in the $xy$-plane is
\[
R_{\varphi}(\theta)R_{\beta+\varphi}(\pi)R_{3\beta+\varphi}(2\pi)R_{\beta+\varphi}(\pi).
\]
Here, $\varphi$ defines the orientation of the rotation axis and
$\beta=\cos^{-1}\left(-\theta/4\pi\right)$. For the present purpose,
we consider the cases $\theta=\pi/2$ and $\pi$, where $\beta_{90}\approx1.7$
and $\beta_{180}\approx1.8$. If the four $\pi$-pulses of this sequence
are separated in time by a delay that is twice the duration of the
initial $\theta$-pulse, this yields a sequence that is not only robust
to amplitude errors, but also to environmental noise: it conforms
to the ``compute, then decouple'' approach and the four $\pi$-pulse
generate the DD cycle.

A robust $\pi$-pulse can also be generated by concatenating 5 $\pi$-pulses
with the phase $\pi/6$, 0, $\pi/2$, 0, $\pi/6$ \cite{985} (see
figure \ref{fig:Different-versions-of}f). This generates an inversion
of the $z$-component, but in addition also a $-\pi/3$ rotation around
the $z$-axis. We will refer to this pulse as KDD-5 = $R_{z}(-\pi/3)R_{0}$
= $R_{0}R_{z}(\pi/3)$, where $R_{0}$ is the targeted $\pi$-rotation
around the $x$-axis. If it is used in a sequence like randomized
benchmarking, the additional $z-$rotation can be taken into account
by adjusting the coordinate system. 

To determine the average fidelity of the gate operations, we use the
scheme proposed by \cite{PhysRevA.77.012307,Ryan:2009fk}. It requires
the application of $\pi/2$ and $\pi$ rotations around the axes of
the coordinate system. The individual gates are 1-qubit Clifford gates
$C=PG$, where $P$ indicates a unit operation or a $\pi$-rotation
around one of the coordinate axes (8 different operations) and $G$
a $\pi/2$ rotation around a coordinate axis (6 different operations,
48 total). Rotations around $\pm z$ axes are implemented through
a change to the rotating frame definition \cite{Ryan:2009fk,PhysRevA.77.012307}.
At the end of the sequence a recovery operation is applied that corresponds
to the inverse of the sequence up to this point, so that the full
sequence becomes a unit operation in the absence of imperfections.
For optimal randomization, the recovery operation itself consists
of two random $P$ operations sandwiching the complement $R$, which
is another Clifford gate. For a single qubit, the effect of a non-ideal
identity operation can be written as
\[
\rho_{out}=(1-d)\rho_{in}+\frac{d}{2}\mathbf{1}.
\]

Here, $\rho_{in}$ is the initial state, $\mathbf{1}$ the unit operator
and $d$ is the depolarizing parameter. Here, we use the fact that
for a sequence of Clifford gates, the result can be represented as
a depolarization channel \cite{pra012304,JOptB7S347}. The trace fidelity
is then
\[
F=tr\{\rho_{out}\rho_{in}\}=1-\frac{d}{2}.
\]

and the error per gate is the difference 
\[
EPG=1-F=\frac{d}{2}.
\]

If the error per gate is purely random, we expect that the average
fidelity for sequences with $m$ gate operations decays as 
\[
F=\frac{1}{2}\left(1+(1-d)^{m}\right)\approx\frac{1}{2}(1+e^{-md}).
\]

For the experiment, we used the $^{13}$C nuclear spins of adamantane,
a molecular crystal. The protons in the same crystal provide the noisy
environment, which leads to a dephasing time of 360 \textmu s. A single
refocusing pulse (Hahn echo) can extend the dephasing time to 740
\textmu s. The ultimate limit on the information storage in this system
is given by the energy relaxation time $T_{1}$=1.52 s.

The system was initially in thermal equilibrium, with $\rho_{in}\propto S_{z}$.
For each measurement, we averaged over a set of 32 random sequences
composed of 1 to 80 gate operations. After the restore operation,
the signal was measured by applying one additional readout pulse that
converted the remaining $S_{z}$ component of the density operator
into transverse magnetization. The free precession signal was acquired,
Fourier-transformed, and the signal corresponding to the CH$_{2}$
carbon was integrated and used as a measure for the survival probability
$e^{-md}$.

Figure \ref{fig:Different-versions-of} gives an overview over the
different sequences tested in this work. First (not shown in the figure)
we tested gates without DD protection, using two different implementation
for P and G rotations: i) simple rectangular pulses and ii) BB1 pulses,
which compensate errors in the amplitudes of the control fields. In
the second approach (\ref{fig:Different-versions-of}a) we generated
a protected G rotation by separating the four $\pi$-pulses of the
BB1 composite rotation by a delay that is twice the duration of the
initial $\pi/2$-pulse. The P rotation in this scheme was implemented
by one KDD-5 pulse. The third set of sequences (\ref{fig:Different-versions-of}b-e)
was designed to reduce experimental imperfections and decouple the
system from its environment by interleaing the P and G rotations with
dynamical decoupling pulses in different ways, using BB1 pulses for
all PG rotations. In b) and c) the DD pulses were also implemented
as BB1 pulses, while in d) and e) the DD pulses were implemented by
rectangular pulses.

\begin{figure}[h]
\noindent \begin{centering}
\includegraphics[scale=0.5]{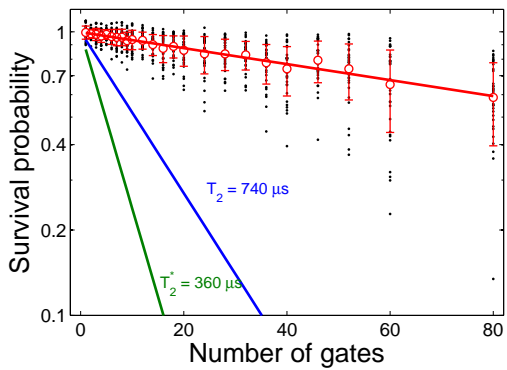}
\par\end{centering}

\protect\caption{Experimental survival probabilities for sequences of BB1-gates as
a function of the number of steps. The circles represent the survival
probability averaged over 32 gate sequences. The solid lines are represent
an exponential fit and theoretical predictions (see text) \label{fig:Experimental-decay-curves}.}
\end{figure}

Figure \ref{fig:Experimental-decay-curves} shows the experimental
results obtained when composite BB1 pulses were used to implement
the Clifford gates. Here, the gate time was $76\,\mathrm{\mu s}$.
The blue and green lines show the theoretical prediction for pure
dephasing with times of $750\,\mathrm{\mu s}$ and $340\,\mathrm{\mu s}$,
respectively, which would result in EPGs of $3.2\%$, and $6.7\%$.
The straight lines through the experimental data points are fits to
exponential decays, the corresponding EPG is $0.32\pm0.03\%$, which
is significantly better then the EPG extrapolated from the $T_{2}$
values. This is a clearly indication that the environment is refocused
during the benchmarking experiment, even without applying DD pulses. 

\begin{figure}
\includegraphics[scale=0.5]{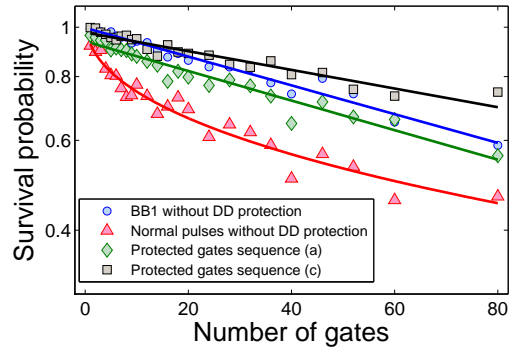} \protect\caption{Experimental results for different type of gates: rectangular pulses
(red), BB1-gates (blue), a compute then decouple approach built from
BB1 pulses and KDD-5 pulses (green), and protected gates by dynamical
decoupling (black). \label{fig:Experimental-decay-curves-2}}
\end{figure}

In Figure \ref{fig:Experimental-decay-curves-2} we compare the BB1
pulses with the three other cases: i) The approach defined in figure
\ref{fig:Different-versions-of}a, ii) interleaving BB1 pulses with
the DD sequence XY-16 (figure \ref{fig:Different-versions-of}c) and
iii) rectangular pulses. The gate time of the of the first scheme
is $88\,\mathrm{\mu s}$ and its performance is comparable to that
of BB1, the observed EPG is $0.34\pm0.03\%$. In the case of interleaving
BB1 pulses with DD, the gate time is $152\,\mathrm{\mu s}$, which
is approximately twice the gate time for BB1 gates without DD protection.
However the observed EPG is $0.22\pm0.03\%$, showing that dynamical
decoupling sequences can further increase the fidelity of quantum
gates. The rectangular pulses do not exhibit an exponential decay,
since those pulses are not built to compensate r.f. field inhomogeneities.
In this case we cannot assign a single value to the EPG. For a quantitative
evaluation, we fitted the experimental data to the function $a^{m^{k}}$.
The parameters obtained from this fit are $a\sim0.88$ and $k\sim0.44$ 

\begin{figure}[h]
\noindent \begin{centering}
\includegraphics[scale=0.5]{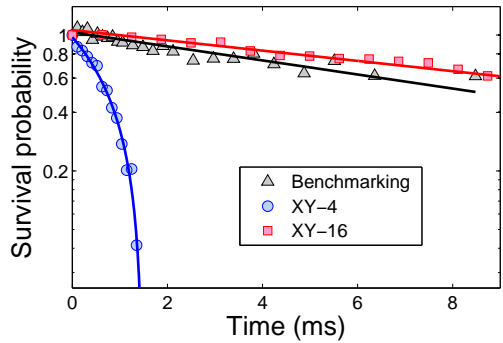}
\par\end{centering}

\protect\caption{Decay of the survival probability measured by interleaving quantum
gates with the dynamical decoupling XY-4 and pure dynamical decoupling,
without gate operations. In the case of the XY-4 sequence, the decay
can be fitted by a quadratic function ($-0.02t{}^{2}-0.64t+0.97$).
The benchmarking and XY-16 results were fitted with the exponential
functions $1.02e^{-t/15.3\,\mathrm{ms}}$ and $1.06e^{-t/16.4\,\mathrm{ms}}$,
respectively\label{fig:Fidelity-decay-of}}
\end{figure}

Figure \ref{fig:Fidelity-decay-of} compares the decay of the survival
probability vs. time for the case of a benchmarking sequence whose
gates are protected against environmental noise by an XY-4 DD sequence
with the decay of coherence if only DD is applied, with no interleaved
gate operations. For the benchmarking experiment the initial state
was thermal equilibrium, with $\rho_{in}\propto S_{z}$, while in
the pure DD experiment the system was initialized in a superposition
state. Clearly, the benchmarking results in much slower decays then
the XY-4 sequence. This result can be traced to the fact that the
pulse imperfections in an XY-4 cycle result in a propagator $U_{4}\approx\mathbf{1}+Q$,
where $Q$ is the error term. In the case of DD, the overall propagator
is $U_{4n}=U_{4}^{n}(\mathbf{1}+Q)^{n}\approx\mathbf{1}+nQ$ : for
small number of cycles $n$, the error grows linearly. In the case
of benchmarking, the gates that are interleaved with the DD sequence
modulate the error in a random manner, resulting in a random walk
in Liouville space. As a result, an increase $\propto n$ may be expected
for a random sequence. This is an expression of the general principle
that in most cases, a decoupling scheme exists that is better than
a simple repetition of the same basic cycle.

The experimental data clearly show that a robust design of the gate
operations can make their performance largely independent of experimental
imperfections. This does not mean that the error per gate can be made
arbitrarily small, since not all interactions that contribute to the
errors are actually suppressed. In particular, two types of environmental
noise cannot be refocused by explicit or implicit DD: (i) interactions
that are not linear in the qubit operators $\vec{S}$ and (ii) interactions
whose correlation time is shorter than the experimentally accessible
time scales. In the present system, $^{13}$C-$^{13}$C couplings
are an example of (i): they are bilinear in the spin interactions
and are not eliminated by the DD sequences used here, although other
sequences exist for suppressing them \cite{6051,Krojanski:2006lr}.
Processes of category (ii) are those that contribute to the energy
relaxation ($T_{1}$-processes): their correlation times must be comparable
to or shorter than the Larmor precession period. Their contributions
can be estimated from available experimental data. The longest dephasing
times that have been reached in this system for isotropic decoupling
sequences indicate that $\tau_{1}\approx50$ ms \cite{Souza13102012}.
This value was achieved by using robust pulses and interpulse delays
close to zero and it is also in qualitative agreement with numerical
estimates for the broadening by $^{13}$C-$^{13}$C dipolar coupling
in natural abundance adamantane. The contribution from fast processes
can be estimated from the energy relaxation time as $\tau_{2}=T_{1}\approx1.52$
s. We can therefore neglect processes of type (ii). This indicates
that further improvements would require pulses that compensate also
for homonuclear (bilinear) interactions. If the remaining dephasing
rate is $T_{2}^{-1}$, this should contribute an error per gate of
$\approx\tau/T_{2}$

\begin{center}
\begin{table}
\begin{centering}
\begin{tabular}{|l|c|c|c|c|c|c|}
\hline 
Gate & BB1 & (a) &  (b) &  (c) & (d) & (e)\tabularnewline
\hline 
\hline 
$\tau${[}\textmu s{]} & 76 & 88 & 116 & 152 & 336 & 384\tabularnewline
\hline 
EPG$_{m}${[}$10^{-4}${]} & 5 & 6 & 8 & 10 & 22 & 25\tabularnewline
\hline 
EPG$_{M}${[}$10^{-4}${]} & 317 & 364 & 472 & 604 & 1191 & 1322\tabularnewline
\hline 
EPG$_{\mathrm{exp}}${[}$10^{-4}${]} & 32$\pm$3 & 34$\pm$3 & 28$\pm$3 & 22$\pm$3 & 172$\pm$6 & 47$\pm$3\tabularnewline
\hline 
\end{tabular}
\par\end{centering}

\protect\caption{Summary of the gate operations tested. For each type of gate, the
duration $\tau$ for a $PG$ operation is given, the limiting error
per gates, EPG$_{M}$ and EPG$_{m}$, which is expected for pure dephasing
with a dephasing time measured by Hahn Echo and pure dynamical decoupling,
respectively. The labels a-e refers to the gates types illustrated
in the figure \ref{fig:Different-versions-of} .\label{tab:Summary-of-the}}
\end{table}

\par\end{center}

Table \ref{tab:Summary-of-the} summarizes the results. For each type
of gate it shows the observed EPG$_{\mathrm{exp}}$. All gate operations
include decoupling elements and thus result in an error per gate that
is lower than the error rate EPG$_{M}$ estimated from the dephasing
time measured by Hahn Echo. This is a clear indication that gate operations
alone provide some dynamical decoupling. Each sequence is also compared
to the minimum EPG$_{m}$ values we could reach in the present system,
EPG$_{m}\approx\tau/T_{2}$, which is expected from pure dephasing
due to the remaining environmental noise. These values were obtained
by DD with robust pulses and very short delays between the pulses\cite{Souza13102012}
. The best performance was achieved by the gate type (c), which corresponds
to BB1 gates protected by the DD sequence XY-16.

The implementation of robust high-fidelity gate operations is an essential
step towards reliable and scalable quantum computing. In this work
we use randomized benchmarking of single qubit quantum gates to compare
the performance of different families of gates that compensate errors
in the control field amplitudes and decouple the system from the environmental
noise. In some cases the total duration of the experiments (from$\sim1$
ms to $\sim30$ ms) exceeds the dephasing time, measured by Hahn echoes
($760$ \textmu s), by almost two orders of magnitude. We could obtain
average fidelities exceeding the expected limit from the dephasing
induced by the environment. This is a clear indication that the effect
of the noisy environment is reduced by the sequence of applied gate
operations. The best average fidelity observed, 99.8\% , was achieved
by interleaving gate operations implemented by composite pulses with
the dynamical decoupling sequence XY-16.

The improvements in gate accuracy by decoupling methods, as observed
in this work, implies a reduction of the overhead cost of QEC since
more noise can be tolerated by QEC codes combined with DD than by
QEC alone\cite{PhysRevA.84.012305}. Therefore, future work will be
devoted to investigate the performance of QEC codes combined with
dynamical decoupling sequences and to test the performance of DD protected
gates in systems with multiple qubits.

This work is suported by CNPq, FAPERJ, the Brazilian National Institute
of Science and Technology for Quantum Information (INCT-IQ) and CAPES
Ciência sem Fronteiras program (Grant 084/2012). 

\bibliographystyle{apsrev}
\bibliography{Benchmarking}

\begin{thebibliography}{25}
\expandafter\ifx\csname natexlab\endcsname\relax\def\natexlab#1{#1}\fi
\expandafter\ifx\csname bibnamefont\endcsname\relax
  \def\bibnamefont#1{#1}\fi
\expandafter\ifx\csname bibfnamefont\endcsname\relax
  \def\bibfnamefont#1{#1}\fi
\expandafter\ifx\csname citenamefont\endcsname\relax
  \def\citenamefont#1{#1}\fi
\expandafter\ifx\csname url\endcsname\relax
  \def\url#1{\texttt{#1}}\fi
\expandafter\ifx\csname urlprefix\endcsname\relax\def\urlprefix{URL }\fi
\providecommand{\bibinfo}[2]{#2}
\providecommand{\eprint}[2][]{\url{#2}}

\bibitem[{\citenamefont{Knill et~al.}(1998)\citenamefont{Knill, Laflamme, and
  Zurek}}]{2711}
\bibinfo{author}{\bibfnamefont{E.}~\bibnamefont{Knill}},
  \bibinfo{author}{\bibfnamefont{R.}~\bibnamefont{Laflamme}}, \bibnamefont{and}
  \bibinfo{author}{\bibfnamefont{W.~H.} \bibnamefont{Zurek}},
  \bibinfo{journal}{Proc. R. Soc. Lond. A} \textbf{\bibinfo{volume}{454}},
  \bibinfo{pages}{365} (\bibinfo{year}{1998}).

\bibitem[{\citenamefont{Preskill}(1998)}]{3921}
\bibinfo{author}{\bibfnamefont{J.}~\bibnamefont{Preskill}},
  \bibinfo{journal}{Proc. R. Soc. Lond. A} \textbf{\bibinfo{volume}{454}},
  \bibinfo{pages}{385} (\bibinfo{year}{1998}).

\bibitem[{\citenamefont{Zhang et~al.}(2014)\citenamefont{Zhang, Souza, Brandao,
  and Suter}}]{PhysRevLett.112.050502}
\bibinfo{author}{\bibfnamefont{J.}~\bibnamefont{Zhang}},
  \bibinfo{author}{\bibfnamefont{A.~M.} \bibnamefont{Souza}},
  \bibinfo{author}{\bibfnamefont{F.~D.} \bibnamefont{Brandao}},
  \bibnamefont{and} \bibinfo{author}{\bibfnamefont{D.}~\bibnamefont{Suter}},
  \bibinfo{journal}{Phys. Rev. Lett.} \textbf{\bibinfo{volume}{112}},
  \bibinfo{pages}{050502} (\bibinfo{year}{2014}).

\bibitem[{\citenamefont{Knill et~al.}(2000)\citenamefont{Knill, Laflamme,
  Martinez, and Tseng}}]{2713}
\bibinfo{author}{\bibfnamefont{E.}~\bibnamefont{Knill}},
  \bibinfo{author}{\bibfnamefont{R.}~\bibnamefont{Laflamme}},
  \bibinfo{author}{\bibfnamefont{R.}~\bibnamefont{Martinez}}, \bibnamefont{and}
  \bibinfo{author}{\bibfnamefont{C.-H.} \bibnamefont{Tseng}},
  \bibinfo{journal}{Nature} \textbf{\bibinfo{volume}{404}},
  \bibinfo{pages}{368} (\bibinfo{year}{2000}).

\bibitem[{\citenamefont{Bravyi and Kitaev}(2005)}]{PhysRevA.71.022316}
\bibinfo{author}{\bibfnamefont{S.}~\bibnamefont{Bravyi}} \bibnamefont{and}
  \bibinfo{author}{\bibfnamefont{A.}~\bibnamefont{Kitaev}},
  \bibinfo{journal}{Phys. Rev. A} \textbf{\bibinfo{volume}{71}},
  \bibinfo{pages}{022316} (\bibinfo{year}{2005}).

\bibitem[{\citenamefont{Souza et~al.}(2011)\citenamefont{Souza, Zhang, Ryan,
  and Laflamme}}]{Souza:2011fk}
\bibinfo{author}{\bibfnamefont{A.~M.} \bibnamefont{Souza}},
  \bibinfo{author}{\bibfnamefont{J.}~\bibnamefont{Zhang}},
  \bibinfo{author}{\bibfnamefont{C.~A.} \bibnamefont{Ryan}}, \bibnamefont{and}
  \bibinfo{author}{\bibfnamefont{R.}~\bibnamefont{Laflamme}},
  \bibinfo{journal}{Nat Commun} \textbf{\bibinfo{volume}{2}}
  (\bibinfo{year}{2011}), \urlprefix\url{http://dx.doi.org/10.1038/ncomms1166}.

\bibitem[{\citenamefont{Zheng et~al.}(2015)\citenamefont{Zheng, Yu, Pan, Zhang,
  Li, Li, Suter, Zhou, Peng, and Du}}]{PhysRevA.91.022314}
\bibinfo{author}{\bibfnamefont{W.}~\bibnamefont{Zheng}},
  \bibinfo{author}{\bibfnamefont{Y.}~\bibnamefont{Yu}},
  \bibinfo{author}{\bibfnamefont{J.}~\bibnamefont{Pan}},
  \bibinfo{author}{\bibfnamefont{J.}~\bibnamefont{Zhang}},
  \bibinfo{author}{\bibfnamefont{J.}~\bibnamefont{Li}},
  \bibinfo{author}{\bibfnamefont{Z.}~\bibnamefont{Li}},
  \bibinfo{author}{\bibfnamefont{D.}~\bibnamefont{Suter}},
  \bibinfo{author}{\bibfnamefont{X.}~\bibnamefont{Zhou}},
  \bibinfo{author}{\bibfnamefont{X.}~\bibnamefont{Peng}}, \bibnamefont{and}
  \bibinfo{author}{\bibfnamefont{J.}~\bibnamefont{Du}}, \bibinfo{journal}{Phys.
  Rev. A} \textbf{\bibinfo{volume}{91}}, \bibinfo{pages}{022314}
  (\bibinfo{year}{2015}),
  \urlprefix\url{http://link.aps.org/doi/10.1103/PhysRevA.91.022314}.

\bibitem[{\citenamefont{Viola et~al.}(1999)\citenamefont{Viola, Knill, and
  Lloyd}}]{5916}
\bibinfo{author}{\bibfnamefont{L.}~\bibnamefont{Viola}},
  \bibinfo{author}{\bibfnamefont{E.}~\bibnamefont{Knill}}, \bibnamefont{and}
  \bibinfo{author}{\bibfnamefont{S.}~\bibnamefont{Lloyd}},
  \bibinfo{journal}{Phys. Rev. Lett.} \textbf{\bibinfo{volume}{82}},
  \bibinfo{pages}{2417} (\bibinfo{year}{1999}).

\bibitem[{\citenamefont{{\'A}lvarez et~al.}(2010)\citenamefont{{\'A}lvarez,
  Ajoy, Peng, and Suter}}]{PhysRevA.82.042306}
\bibinfo{author}{\bibfnamefont{G.~A.} \bibnamefont{{\'A}lvarez}},
  \bibinfo{author}{\bibfnamefont{A.}~\bibnamefont{Ajoy}},
  \bibinfo{author}{\bibfnamefont{X.}~\bibnamefont{Peng}}, \bibnamefont{and}
  \bibinfo{author}{\bibfnamefont{D.}~\bibnamefont{Suter}},
  \bibinfo{journal}{Phys. Rev. A} \textbf{\bibinfo{volume}{82}},
  \bibinfo{pages}{042306} (\bibinfo{year}{2010}).

\bibitem[{\citenamefont{Ng et~al.}(2011)\citenamefont{Ng, Lidar, and
  Preskill}}]{PhysRevA.84.012305}
\bibinfo{author}{\bibfnamefont{H.~K.} \bibnamefont{Ng}},
  \bibinfo{author}{\bibfnamefont{D.~A.} \bibnamefont{Lidar}}, \bibnamefont{and}
  \bibinfo{author}{\bibfnamefont{J.}~\bibnamefont{Preskill}},
  \bibinfo{journal}{Phys. Rev. A} \textbf{\bibinfo{volume}{84}},
  \bibinfo{pages}{012305} (\bibinfo{year}{2011}).

\bibitem[{\citenamefont{Khodjasteh and
  Viola}(2009{\natexlab{a}})}]{khodjasteh:080501}
\bibinfo{author}{\bibfnamefont{K.}~\bibnamefont{Khodjasteh}} \bibnamefont{and}
  \bibinfo{author}{\bibfnamefont{L.}~\bibnamefont{Viola}},
  \bibinfo{journal}{Phys. Rev. Lett.} \textbf{\bibinfo{volume}{102}},
  \bibinfo{eid}{080501} (pages~\bibinfo{numpages}{4})
  (\bibinfo{year}{2009}{\natexlab{a}}).

\bibitem[{\citenamefont{Khodjasteh et~al.}(2010)\citenamefont{Khodjasteh,
  Lidar, and Viola}}]{PhysRevLett.104.090501}
\bibinfo{author}{\bibfnamefont{K.}~\bibnamefont{Khodjasteh}},
  \bibinfo{author}{\bibfnamefont{D.~A.} \bibnamefont{Lidar}}, \bibnamefont{and}
  \bibinfo{author}{\bibfnamefont{L.}~\bibnamefont{Viola}},
  \bibinfo{journal}{Phys. Rev. Lett.} \textbf{\bibinfo{volume}{104}},
  \bibinfo{pages}{090501} (\bibinfo{year}{2010}).

\bibitem[{\citenamefont{West et~al.}(2010)\citenamefont{West, Lidar, Fong, and
  Gyure}}]{prl230503}
\bibinfo{author}{\bibfnamefont{J.~R.} \bibnamefont{West}},
  \bibinfo{author}{\bibfnamefont{D.~A.} \bibnamefont{Lidar}},
  \bibinfo{author}{\bibfnamefont{B.~H.} \bibnamefont{Fong}}, \bibnamefont{and}
  \bibinfo{author}{\bibfnamefont{M.~F.} \bibnamefont{Gyure}},
  \bibinfo{journal}{Phys. Rev. Lett.} \textbf{\bibinfo{volume}{105}},
  \bibinfo{pages}{230503} (\bibinfo{year}{2010}).

\bibitem[{\citenamefont{Khodjasteh and
  Viola}(2009{\natexlab{b}})}]{PhysRevA.80.032314}
\bibinfo{author}{\bibfnamefont{K.}~\bibnamefont{Khodjasteh}} \bibnamefont{and}
  \bibinfo{author}{\bibfnamefont{L.}~\bibnamefont{Viola}},
  \bibinfo{journal}{Phys. Rev. A} \textbf{\bibinfo{volume}{80}},
  \bibinfo{pages}{032314} (\bibinfo{year}{2009}{\natexlab{b}}).

\bibitem[{\citenamefont{Cappellaro et~al.}(2009)\citenamefont{Cappellaro,
  Jiang, Hodges, and Lukin}}]{PhysRevLett.102.210502}
\bibinfo{author}{\bibfnamefont{P.}~\bibnamefont{Cappellaro}},
  \bibinfo{author}{\bibfnamefont{L.}~\bibnamefont{Jiang}},
  \bibinfo{author}{\bibfnamefont{J.~S.} \bibnamefont{Hodges}},
  \bibnamefont{and} \bibinfo{author}{\bibfnamefont{M.~D.} \bibnamefont{Lukin}},
  \bibinfo{journal}{Phys. Rev. Lett.} \textbf{\bibinfo{volume}{102}},
  \bibinfo{pages}{210502} (\bibinfo{year}{2009}).

\bibitem[{\citenamefont{Souza et~al.}(2012{\natexlab{a}})\citenamefont{Souza,
  \'Alvarez, and Suter}}]{PhysRevA.86.050301}
\bibinfo{author}{\bibfnamefont{A.~M.} \bibnamefont{Souza}},
  \bibinfo{author}{\bibfnamefont{G.~A.} \bibnamefont{\'Alvarez}},
  \bibnamefont{and} \bibinfo{author}{\bibfnamefont{D.}~\bibnamefont{Suter}},
  \bibinfo{journal}{Phys. Rev. A} \textbf{\bibinfo{volume}{86}},
  \bibinfo{pages}{050301} (\bibinfo{year}{2012}{\natexlab{a}}),
  \urlprefix\url{http://link.aps.org/doi/10.1103/PhysRevA.86.050301}.

\bibitem[{\citenamefont{Wimperis}(1994)}]{Wimperis:1994fk}
\bibinfo{author}{\bibfnamefont{S.}~\bibnamefont{Wimperis}},
  \bibinfo{journal}{Journal of Magnetic Resonance, Series A}
  \textbf{\bibinfo{volume}{109}}, \bibinfo{pages}{221} (\bibinfo{year}{1994}).

\bibitem[{\citenamefont{Cho et~al.}(1986)\citenamefont{Cho, Tycko, Pines, and
  Guckenheimer}}]{985}
\bibinfo{author}{\bibfnamefont{H.~M.} \bibnamefont{Cho}},
  \bibinfo{author}{\bibfnamefont{R.}~\bibnamefont{Tycko}},
  \bibinfo{author}{\bibfnamefont{A.}~\bibnamefont{Pines}}, \bibnamefont{and}
  \bibinfo{author}{\bibfnamefont{J.}~\bibnamefont{Guckenheimer}},
  \bibinfo{journal}{Phys. Rev. Lett.} \textbf{\bibinfo{volume}{56}},
  \bibinfo{pages}{1905} (\bibinfo{year}{1986}).

\bibitem[{\citenamefont{Knill et~al.}(2008)\citenamefont{Knill, Leibfried,
  Reichle, Britton, Blakestad, Jost, Langer, Ozeri, Seidelin, and
  Wineland}}]{PhysRevA.77.012307}
\bibinfo{author}{\bibfnamefont{E.}~\bibnamefont{Knill}},
  \bibinfo{author}{\bibfnamefont{D.}~\bibnamefont{Leibfried}},
  \bibinfo{author}{\bibfnamefont{R.}~\bibnamefont{Reichle}},
  \bibinfo{author}{\bibfnamefont{J.}~\bibnamefont{Britton}},
  \bibinfo{author}{\bibfnamefont{R.~B.} \bibnamefont{Blakestad}},
  \bibinfo{author}{\bibfnamefont{J.~D.} \bibnamefont{Jost}},
  \bibinfo{author}{\bibfnamefont{C.}~\bibnamefont{Langer}},
  \bibinfo{author}{\bibfnamefont{R.}~\bibnamefont{Ozeri}},
  \bibinfo{author}{\bibfnamefont{S.}~\bibnamefont{Seidelin}}, \bibnamefont{and}
  \bibinfo{author}{\bibfnamefont{D.~J.} \bibnamefont{Wineland}},
  \bibinfo{journal}{Phys. Rev. A} \textbf{\bibinfo{volume}{77}},
  \bibinfo{pages}{012307} (\bibinfo{year}{2008}).

\bibitem[{\citenamefont{Ryan et~al.}(2009)\citenamefont{Ryan, Laforest, and
  Laflamme}}]{Ryan:2009fk}
\bibinfo{author}{\bibfnamefont{C.~A.} \bibnamefont{Ryan}},
  \bibinfo{author}{\bibfnamefont{M.}~\bibnamefont{Laforest}}, \bibnamefont{and}
  \bibinfo{author}{\bibfnamefont{R.}~\bibnamefont{Laflamme}},
  \bibinfo{journal}{New Journal of Physics} \textbf{\bibinfo{volume}{11}},
  \bibinfo{pages}{013034} (\bibinfo{year}{2009}).

\bibitem[{\citenamefont{C et~al.}(2009)\citenamefont{C, R, J, and
  E}}]{pra012304}
\bibinfo{author}{\bibfnamefont{D.}~\bibnamefont{C}},
  \bibinfo{author}{\bibfnamefont{C.}~\bibnamefont{R}},
  \bibinfo{author}{\bibfnamefont{E.}~\bibnamefont{J}}, \bibnamefont{and}
  \bibinfo{author}{\bibfnamefont{L.}~\bibnamefont{E}},
  \bibinfo{journal}{Physical Review A} \textbf{\bibinfo{volume}{80}},
  \bibinfo{pages}{012304} (\bibinfo{year}{2009}).

\bibitem[{\citenamefont{J et~al.}(2005)\citenamefont{J, R, and K}}]{JOptB7S347}
\bibinfo{author}{\bibfnamefont{E.}~\bibnamefont{J}},
  \bibinfo{author}{\bibfnamefont{A.}~\bibnamefont{R}}, \bibnamefont{and}
  \bibinfo{author}{\bibfnamefont{Z.}~\bibnamefont{K}}, \bibinfo{journal}{J.
  Opt. B: Quantum Semiclass. Opt.} \textbf{\bibinfo{volume}{7}},
  \bibinfo{pages}{S347} (\bibinfo{year}{2005}).

\bibitem[{\citenamefont{Haeberlen}(1976)}]{6051}
\bibinfo{author}{\bibfnamefont{U.}~\bibnamefont{Haeberlen}}, in
  \emph{\bibinfo{booktitle}{Advances in magnetic resonance}}, edited by
  \bibinfo{editor}{\bibfnamefont{J.}~\bibnamefont{Waugh}}
  (\bibinfo{publisher}{Acad. Press, New York}, \bibinfo{year}{1976}), vol.
  \bibinfo{volume}{Supplement 1} of \emph{\bibinfo{series}{Advances in magnetic
  resonance}}, chap. \bibinfo{chapter}{Supplement 1}.

\bibitem[{\citenamefont{Krojanski and Suter}(2006)}]{Krojanski:2006lr}
\bibinfo{author}{\bibfnamefont{H.~G.} \bibnamefont{Krojanski}}
  \bibnamefont{and} \bibinfo{author}{\bibfnamefont{D.}~\bibnamefont{Suter}},
  \bibinfo{journal}{Phys. Rev. Lett.} \textbf{\bibinfo{volume}{97}},
  \bibinfo{pages}{150503} (\bibinfo{year}{2006}).

\bibitem[{\citenamefont{Souza et~al.}(2012{\natexlab{b}})\citenamefont{Souza,
  {\'A}lvarez, and Suter}}]{Souza13102012}
\bibinfo{author}{\bibfnamefont{A.~M.} \bibnamefont{Souza}},
  \bibinfo{author}{\bibfnamefont{G.~A.} \bibnamefont{{\'A}lvarez}},
  \bibnamefont{and} \bibinfo{author}{\bibfnamefont{D.}~\bibnamefont{Suter}},
  \bibinfo{journal}{Philosophical Transactions of the Royal Society A:
  Mathematical, Physical and Engineering Sciences}
  \textbf{\bibinfo{volume}{370}}, \bibinfo{pages}{4748}
  (\bibinfo{year}{2012}{\natexlab{b}}).

\end{thebibliography}

\end{document}